\documentclass[twocolumn,final,twoside,journal]{IEEEtran}
\usepackage{epsfig,amsfonts,subfigure,color,amsbsy}
\usepackage[nolist]{acronym}
\usepackage{graphicx,cite,amssymb,amsmath,perso,amsthm}
\usepackage{tikz}
\usepackage{multirow}
\usepackage{bigstrut}
\usepackage{psfrag}
\usetikzlibrary{shapes,arrows}
\usepackage{xcolor}
\usepackage{mathtools}
\usepackage{tabularx,array}
\mathtoolsset{showonlyrefs}

\IEEEoverridecommandlockouts

\newcommand{\todo}[1]{}

\begin{document}
\newtheorem{thm}{Theorem}
\newtheorem{lemma}{Lemma}
\newtheorem{definition}{Definition}
\newtheorem{example}{Example}

\newcommand{\load}{\mathsf{G}}
\newcommand{\tp}{\mathsf{T}}
\newcommand{\rate}{\mathsf{R}}
\newcommand{\ind}{d}
\newcommand{\ld}{\lambda}
\newcommand{\EBdegDist}{\left\{\ld_\ind \right\}}
\newcommand{\Ld}{\Lambda}
\newcommand{\BNdegDist}{\left\{\Ld_\ind \right\}}
\newcommand{\davg}{\bar{\mathrm{d}}}
\newcommand{\dmax}{\mathrm{d}_{\mathrm{max}}}
\newcommand{\rd}{\rho}
\newcommand{\ESdegDist}{\left\{\rd_\slotind \right\}}
\newcommand{\Rd}{\mathrm{P}}
\newcommand{\SNdegDist}{\left\{\Rd_\slotind \right\}}
\newcommand{\thr}{\load^{\star}}
\newcommand{\area}{\msr{A}}
\newcommand{\areab}{\area_{\mathsf b}}
\newcommand{\areas}{\area_{\mathsf s}}
\newcommand{\exitb}{\mathsf{f}_{\mathsf b}}
\newcommand{\exits}{\mathsf{f}_{\mathsf s}}
\newcommand{\plos}{\epsilon}
\newcommand{\x}{\mathtt{x}}
\newcommand{\p}{\mathtt{p}}
\newcommand{\q}{\mathtt{q}}
\newcommand{\qtarget}{\bar{\mathtt{q}}}
\newcommand{\plostarget}{\bar{\plos}}
\newcommand{\snr}{\mathsf{b}}
\newcommand{\snravg}{\bar{\mathsf{B}}} %Enrico
\newcommand{\snrthr}{\mathsf{b}^{\star}}
\newcommand{\snrrv}{\mathsf{B}} % Enrico
\newcommand{\power}{\mathsf{p}}
\newcommand{\poweravg}{\bar{\mathsf{P}}}
\newcommand{\powerrv}{\mathsf{P}}
\newcommand{\noise}{\mathsf{N}}
\newcommand{\inter}{\mathsf{I}}
\newcommand{\snrdist}{f_{\snrrv}}
\newcommand{\powerdist}{f_{\powerrv}}
\newcommand{\fail}{\mathsf{F}}
\newcommand{\de}{\mathrm{d}}
\newcommand{\zt}{z_t}
\newcommand{\slotind}{c} %Enrico
\newcommand{\calB}{\mathcal{B}} %Enrico
\newcommand{\calS}{\mathcal{S}} %Enrico
\newcommand{\PLR}{\mathsf{PLR}} %Enrico
\newcommand{\PLRtarget}{\overline{\PLR}} %Enrico
\newcommand{\nslot}{n}
\newcommand{\nuser}{m}

\newcommand{\oleq}[1]{\overset{\text{(#1)}}{\leq}}
\newcommand{\oeq}[1]{\overset{\text{(#1)}}{=}}
\newcommand{\ogeq}[1]{\overset{\text{(#1)}}{\geq}}
\newcommand{\ogeql}[2]{\overset{#1}{\underset{#2}{\gtreqless}}}

%%%%% Color for Editorial Changes
\definecolor{gl}{rgb}{0.0,0.5,0.8}
\definecolor{fc}{rgb}{0.8,0.5,0}
\definecolor{al}{rgb}{1,0.3,0.3}
\newcommand{\giangio}{\textcolor{gl}}
\newcommand{\fede}{\textcolor{fc}}
\newcommand{\alert}{\textcolor{al}}

%%%%%%%%%%%%%%%%%%%%%%%%%%%%%%%%%%%%%%%%%%%%%%%%%%%%%%%%%%%%%%%%%%%%%%%

\title{Irregular Repetition Slotted ALOHA over the Rayleigh Block Fading Channel with Capture}

\author{
    \IEEEauthorblockN{Federico Clazzer\IEEEauthorrefmark{1}, Enrico Paolini\IEEEauthorrefmark{2}, Iacopo Mambelli\IEEEauthorrefmark{3}, {\v C}edomir Stefanovi{\'c}\IEEEauthorrefmark{4}}\\
    \IEEEauthorblockA{\IEEEauthorrefmark{1}Institute of Communications and Navigation of DLR (German Aerospace Center),
    Wessling, Germany.}\\ %\\ Email: Federico.Clazzer@dlr.de}\\
    \IEEEauthorblockA{\IEEEauthorrefmark{2}Department of Electrical, Electronic, and Information Engineering, University of Bologna, Cesena, Italy.}\\%\\ Email: e.paolini@unibo.it}\\
    \IEEEauthorblockA{\IEEEauthorrefmark{3}Fores Engineering S.R.L., Forl{\`i}, Italy.}\\ %Email: iacopo.mambelli@fores.it}\\
    \IEEEauthorblockA{\IEEEauthorrefmark{4} Department of Electronic Systems, Aalborg University, Aalborg, Denmark.}% Email: cs@es.aau.dk}
}

 \maketitle
%%%%%%%%%%%%%%%%%%%%%%%%%%%%%%%%%%%%%%%%%%%%%%%%%%%%%%%%%%%%%%%%%%%%%%%%%%%%%%%%%%%%%%%%%%%%%%%%%%%%
%%%%%

\thispagestyle{empty}

%%%%%%%%%%%%%%%%%%%%%%%%%%%%%%%%%%%%%%%%%%%%%%%%%%%%%%%%%%%%%%%%%%%%%%%%%%%%%%%%%%%%%%%%%%%%%%%%%%%%
%%%%%%%%%%%%%%%%%%%%%%%%%

\begin{abstract}
Random access protocols relying on the transmission of packet replicas in multiple slots and exploiting interference cancellation at the receiver have been shown to achieve performance competitive with that of orthogonal schemes.
%The use of a variable number of copies for each transmitted packet leads to noticeable improvements w.r.t. \ac{SA}.
So far the optimization of the repetition degree profile, defining the probability for a user to transmit a given number of replicas, has mainly been performed targeting the collision channel model. In this paper the analysis is extended to a block fading channel model, also assuming capture effect at the receiver. Density evolution equations are developed for the new setting and, based on them, some repetition degree profiles are optimized and analyzed via Monte Carlo simulation in a finite frame length setting. The derived distributions are shown to achieve throughputs largely exceeding $\mathbf{1}$ $\mathrm{\mathbf{[packet/slot]}}$.
\end{abstract}

\thispagestyle{empty}
\setcounter{page}{1}

\thispagestyle{empty}
\pagestyle{empty}

%%%%%%%%%%%%%%%%%%%%%%%%%%%%%%%%%%%%%%%%%%%%%%%%%%%%%%%%%%%%%%%%%%%%%%%
%\input{./tex/displaynotation}t
\begin{acronym}
\acro{BEC}{binary erasure channel}
\acro{BP}{belief propagation}
\acro{CRDSA}{contention resolution diversity slotted Aloha}
\acro{DE}{density evolution}
\acro{EXIT}{extrinsic information transfer}
\acro{i.i.d.}{independent and identically distributed}
\acro{IC}{interference cancellation}
\acro{IRSA}{irregular repetition slotted Aloha}
\acro{LDPC}{low-density parity-check}
\acro{LHS}{left-hand side}
\acro{LT}{Luby-transform}
\acro{MAC}{medium access control}
\acro{MAP}{maximum a posteriori probability}
\acro{PLR}{packet loss rate}
\acro{r.v.}{random variable}
\acro{RHS}{right-hand side}
\acro{SINR}{signal-to-interference-and-noise ratio}
\acro{SNR}{signal-to-noise ratio}
\acro{p.m.f.}{probability mass function} %Enrico
\acro{SIC}{successive interference cancellation} %Enrico
\acro{SA}{slotted Aloha} %Enrico
\end{acronym}

%%%%%%%%%%%%%%%%%%%%%%%%%%%%%%%%%%%%%%%%%%%%%%%%%%%%%%%%%%%%%%%%%%%%%%%

%\clearpage
%\tableofcontents
%\clearpage

%%%%%%%%%%%%%%%%%%%%%%%%%%%%%%%%%%%%%%%%%%%%%%%%%%%%%%%%%%%%%%%%%%%%%%%

\section{Introduction}\label{sec:Intro}

Framed slotted ALOHA (FSA) \cite{OIN1977} is a random access scheme in which link-time is divided in frames and frames consist of slots. 
Users are frame- and slot-synchronized; a user contends by transmitting its packet in a randomly chosen slot of the frame.
For the collision channel model, the maximum value of the expected asymptotic throughput of FSA is $1/e \, \mathrm{[packet/slot]}$.

A substantial throughput gain can be achieved by modifying FSA such that (i) the users transmit replicas of their packets in several slots chosen at random, where each replica embeds pointers to slots containing the other replicas, and (ii) when a replica has been decoded at the receiver, all the other replicas are removed through the use of \ac{IC} \cite{DeGaudenzi07:CRDSA, Liva11:IRSA}.
Specifically, it has been shown that this enhanced version of FSA, named \ac{IRSA} \cite{Liva11:IRSA}, bears striking similarities to the erasure-correcting coding framework in which the decoding is performed with iterative belief-propagation.
These insights were followed by a strand of works applying various concepts and tools of erasure-coding theory to design slotted ALOHA-based schemes, commonly referred to as coded slotted ALOHA.\footnote{We refer the interested reader to \cite{PSLP2014} for an overview.}
The overall conclusion is that coded slotted ALOHA schemes can asymptotically achieve the expected throughput of $1$ $\mathrm{[packet/slot]}$, which is the ultimate limit for the collision channel model.

On the other hand, the collision channel model is a rather simple one, which assumes that (i) noise can be neglected, such that a transmission can be decoded from a singleton slot by default, and (ii) no transmission can be decoded from a collision slot.
This model has a limited practical applicability and does not describe adequately the wireless transmission scenarios where the impact of fading and noise cannot be neglected.
In particular, fading may incur power variations among signals observed in collisions slot, allowing for the \emph{capture effect} to occur, when sufficiently strong signals may be decoded.
In the context of slotted ALOHA, numerous works assessed the performance of the scheme for different capture effect models \cite{R1975,N1984,WEM1989,ZR1994,Z1997,ZZ2012}.
One of the standardly used models is the threshold-based one, in which a packet is captured, i.e., decoded, if its \ac{SINR} is higher then a predefined threshold, c.f. \cite{ZR1994,Z1997,NEW2007,ZZ2012}.

A brief treatment of the capture effect in \ac{IRSA} framework was made in \cite{Liva11:IRSA}, pointing out the implications related to the asymptotic analysis.
In \cite{SMP2014}, the method for the computation of capture probabilities for the threshold-based model in single-user detection systems with Rayleigh fading was presented and instantiated for the frameless ALOHA framework \cite{Stefanovic12:FramelessAloha}.

In this paper, we extend the treatment of the threshold-based capture effect for \ac{IRSA} framework.
First, we derive the exact expressions of capture probabilities for the threshold-based model and Rayleigh block-fading channel.
Next we formulate the asymptotic performance analysis.
We then optimize the scheme, in terms of deriving the optimal repetition strategies that maximize throughput given a target \ac{PLR}. %\CS{we should also mention here thresholds and average degree.}\FC{Shall we, or can we leave them out sparing one sentence?}
Finally, the obtained distributions are investigated in the finite frame length scenario via simulations.
We show that \ac{IRSA} exhibits a remarkable throughput performance that is well over $1$ $\mathrm{[packet/slot]}$, for target \ac{PLR}, SNR and threshold values that are valid in practical scenarios.
This is demonstrated both for asymptotic and finite frame length cases, showing also that the finite-length performance indeed tends to the asymptotic one as the frame length increases.

The paper is organized as follows.
Section~\ref{sec:Preliminaries} introduces the system model.
Section~\ref{sec:decoding_probs} deals with derivation of capture probabilities for Rayleigh block-fading channels, which are used in Section~\ref{sec:DE} to evaluate the asymptotic performance of IRSA.
The study of optimal repetition strategies is done in Section~\ref{sec:Performance}, including their characterization in the finite-frame length case.
Section~\ref{sec:Conclusions} concludes the paper.

\section{System Model}\label{sec:Preliminaries}

\subsection{Access Protocol}

For the sake of simplicity, we focus on a single batch arrival of $\nuser$ users %, indexed by $m \in \{1,2,\dots,\nuser\}$,
each having a single packet (or \emph{burst}) and contending for the access to the common receiver. The link time is organized in a \ac{MAC} frame of duration $T_F$, divided into $\nslot$ slots of equal duration $T_S=T_F / \nslot$, indexed by $j \in \{1, 2, \dots,  \nslot\}$. The transmission time of each packet equals the slot duration. The system load $\load$ is defined as
\begin{align}
\load = \frac{\nuser}{\nslot}\,\,\,\, [\mathrm{packet}/\mathrm{slot}] \, .
\end{align}

According to the \ac{IRSA} protocol, each user selects a repetition degree $\ind$ by sampling a \ac{p.m.f.} $\BNdegDist_{\ind=2}^{\dmax}$ and transmits $\ind$ identical replicas of its burst in $d$ randomly chosen slots of the frame. It is assumed that the header of each burst replica carries information about the locations (i.e., slot indexes) of all $\ind$ replicas. The \ac{p.m.f.} $\BNdegDist$ is the same for all users and is sampled independently by different users, in an uncoordinated fashion.
The average burst repetition degree is $\davg =\sum_{\ind=2}^{\dmax} \ind \, \Ld_\ind$, and its inverse
\begin{align}\label{eq:rate}
\rate = \frac{1}{\davg}
\end{align}
is called the \emph{rate} of the IRSA scheme. %\CS{The quantity $10 \log_{10} (1/\rate)$ represents the increment (in dB) of energy per burst with respect to the case of absence of burst repetitions.}\FC{I would remove it, unless there is a reason why to keep it. It is of course correct, but we do not need it for our contribution I think.}
Each user is then unaware of the repetition degree employed by the other users contending for the access.
The number of burst replicas colliding in slot $j$ is denoted by $c_j \in \{1,2,\dots,\nuser\}$. Burst replicas colliding in slot $j$ are indexed by $i \in \{1,2,\dots,c_j\}$.

\subsection{Received Power and Fading Models}

We consider a Rayleigh block fading channel model, i.e., fading is Rayleigh distributed, constant and frequency flat in each block, while it is \ac{i.i.d.} on different blocks. Independent fading between different burst replicas is also assumed. In this way, the power of a burst replica $i \in \{1,2,\dots,c_j\}$ received in slot $j$, denoted as $\powerrv_{ij}$, is modeled as a \ac{r.v.} with negative exponential distribution
\[
\powerdist\left(\power\right)=
\begin{cases}
\frac{1}{\poweravg} \exp \left[-\frac{\power}{\poweravg} \right] , & \power\geq 0 \\
0, & \text{otherwise}
\end{cases}
\]
where $\poweravg$ is the average received power. This is assumed to be the same for all burst replicas received in the \ac{MAC} frame by using, e.g., a long-term power control. The \acp{r.v.} $\powerrv_{ij}$ are \ac{i.i.d.}  for all $(i,j)$ pairs. If we denote by $\noise$ the noise power, the \ac{SNR} \ac{r.v.} $\snrrv_{ij} = \powerrv_{ij}/\noise$ is also exponentially distributed as
\[
\snrdist\left(\snr\right)=
\begin{cases}
\frac{1}{\snravg}\exp \left[ -\frac{\snr}{\snravg} \right], & \snr \geq 0 \\
0, & \text{otherwise}
\end{cases}
\]
where the average \ac{SNR} is given by
\[
\snravg=\frac{\poweravg}{\noise}.
\]

\subsection{Graph Representation}\label{subsec:graph}

\begin{figure}[tb]
\begin{center}
\includegraphics[width=0.75\columnwidth,draft=false]{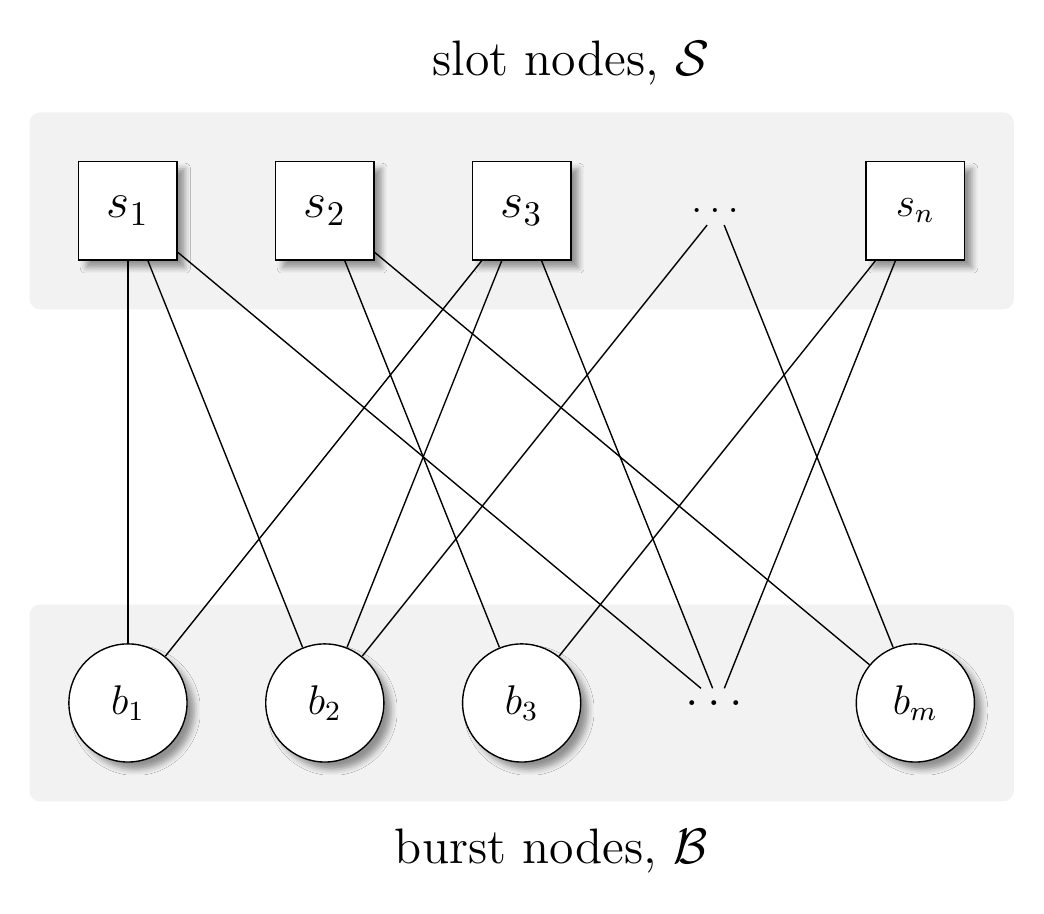}
\end{center}
\caption{Graph representation of MAC frame.}\label{fig:graph}
\end{figure}

In order to analyze the \ac{SIC} process, we introduce the graph representation of a \ac{MAC} frame\cite{Liva11:IRSA}. As depicted in Fig.~\ref{fig:graph}, a \ac{MAC} frame is represented as a bipartite graph $\mathcal{G}=(\calB,\calS,E)$ consisting of a set $\calB$ of $\nuser$ burst nodes (or user nodes), one for each user, a set $\calS$ of $\nslot$ slot nodes, one per slot, and a set $E$ of edges, one per transmitted burst replica. A burst node $b_k\in \calB$ is connected to a slot node $s_j\in \calS$ if and only if user $k$ has a burst replica sent in the $j$-th slot of the frame. The \emph{node degree} represents the number of edges emanating from a node.

For the upcoming analysis it is convenient to resort to the concept of \emph{node-} and \emph{edge-perspective degree distributions}. The burst node degree distribution from a node perspective is identified by the above-defined \ac{p.m.f.} $\BNdegDist_{\ind=2}^{\dmax}$. Similarly, the slot node degree distribution from a node perspective is defined as $\SNdegDist_{\slotind=0}^\nuser$, where $\Rd_\slotind$ is the probability that a slot node has $\slotind$ connections (i.e., that $\slotind$ burst replicas have been received in the corresponding slot). The probability $\Rd_\slotind$ may be easily calculated by observing that $(\load / \rate) / \nuser$ is the probability that the generic user transmits a burst replica in a specific slot. Since users behave independently of each other, we obtain
\begin{align}
\Rd_\slotind={\nuser \choose \slotind}
\left(\frac{\load/\rate}{\nuser}\right)^{\slotind} \left(1-\frac{\load/\rate}{\nuser}\right)^{\nuser - \slotind}\, .
\end{align}

The polynomial representations for both node-perspective degree distributions are given by
\begin{align}
\Ld(x) = \sum_{\ind=2}^{\dmax} \Ld_\ind \, x^\ind \quad \mathrm{and} \quad \Rd(x) = \sum_{\slotind=0}^\nuser \Rd_\slotind \, x^\slotind.
\end{align}
For $\nuser \rightarrow \infty$ and constant $ \load / \rate$, $\Rd(x) = \exp \left\{ -\frac{\load}{\rate}(1-x) \right\}$.
Degree distributions can also be defined from an edge-perspective. Adopting a notation similar to the one used for the node-perspective distributions, we define the edge-perspective burst node degree distribution as the \ac{p.m.f.} $\EBdegDist_{\ind=2}^{\dmax}$, where $\ld_\ind$ is the probability that a given edge is connected to a burst node of degree $\ind$. Likewise, we define the edge-perspective slot node degree distribution as the \ac{p.m.f.} $\ESdegDist_{\slotind=0}^\nuser$, where $\rd_\slotind$ is the probability that an edge is connected to a slot node of degree $\slotind$. From the definitions we have $\ld_\ind = \ind \, \Ld_\ind / \left(\sum_t t \, \Ld_t \right)$ and $\rd_\slotind = \slotind \, \Rd_\slotind / \left(\sum_t t \, \Rd_t \right)$;
it can be shown that, for $\nuser \rightarrow \infty$ and constant $ \load / \rate$,  $ \rd_\slotind  = \exp\{-\load / \rate\} (\load / \rate)^{\slotind-1} / (\slotind-1)! $.
The corresponding polynomial representation are
$\ld(x) = \sum_{\ind=2}^{\dmax} \ld_\ind \, x^{\ind-1}$ and $\rd(x) = \sum_{\slotind=0}^\nuser \rd_\slotind \, x^{\slotind-1}$.
%\begin{align}
%\ld(x) = \sum_{\ind=2}^{\dmax} \ld_\ind \, x^{\ind-1} \quad \mathrm{and} \quad \rd(x) = \sum_{\slotind=0}^\nuser \rd_\slotind \, x^{\slotind-1} \, .
%\end{align}
%
%and
%%
%$$
%\rd(x) = \sum_{\slotind=0}^\nuser \rd_\slotind \, x^{\slotind-1}
%$$
%
Note that $\ld(x)=\Ld'(x)/\Ld'(1)$ and $\rd(x)=\Rd'(x)/\Rd'(1)$.\footnote{Notation $f'(x)$ denotes the derivative of $f(x)$.}

%\textcolor{red}{Recalling the definition \eqref{eq:rate} of rate, we may write}
%%
%\begin{equation}
%\label{rho_d}
%\textcolor{red}{\rd_d = e^{-\frac{\load}{\rate}}\frac{\left(\frac{\load}{\rate}\right)^{d-1}}{(d-1)!} \, .}
%\end{equation}

\subsection{Receiver Operation}
\label{subsec:receiver}

In our model, the receiver is always able to detect burst replicas received in a slot, i.e., to discriminate between an empty slot where only noise samples are present and a slot in which at least one burst replica has been received. We assume that the receiver is able to obtain perfect channel state information, also in the slots undergoing packet collisions. Moreover, a threshold-based capture model for the receiver is assumed, by which the generic burst replica $i$ is successfully decoded (i.e., captured) in slot $j$ if the \ac{SINR} exceeds a certain threshold $\snrthr$, namely,
\begin{align}\label{eq:capture_model}
\Pr \{ \text{burst replica $i$ decoded} \} =
\begin{cases}
1, & \frac{\powerrv_{ij}}{\noise + \inter_{ij}} \geq \snrthr \\
0, & \text{otherwise}.
\end{cases}
\end{align}
The quantity $\inter_{ij}$ in \eqref{eq:capture_model} denotes the power of the interference impairing replica $i$ in slot $j$. In our system model the threshold $\snrthr$ fulfills $\snrthr \geq 1$, which corresponds to a conventional narrowband single-antenna system. As we are considering a \ac{SIC}-based receiver, under the assumption of perfect \ac{IC} $\inter_{ij}$ is equal to sum of the powers of those bursts that have not yet been cancelled from slot $j$ in previous iterations (apart form burst $i$). %
%and of the residual power of the bursts that have already been cancelled.
Specifically,
\begin{align}\label{eq:interference_expression}
\inter_{ij} = \sum_{u \in \mathcal{R}_j \setminus \{ i \} } \powerrv_{uj} %+ \sum_{v \in \mathcal{C}_j} Q_{vj}
\end{align}
where $\powerrv_{uj}$ is the power of burst replica $u$ not yet cancelled in slot $j$ and $\mathcal{R}_j$ denotes the set of remaining burst replicas in slot $j$. %
%, $Q_{vj}$ is the residual power of burst replica $v$ already cancelled from slot $j$, and $\mathcal{C}_j$ denotes the set of all burst replicas so far cancelled from slot $j$. Clearly we have $\mathcal R_j \cup \, \mathcal C_j = \{1,2,\dots,c_j\}$ and $\mathcal R_j \cap \mathcal C_j = \emptyset$.
%In case of imperfect \ac{IC}, the residual powers $ Q_{vj} $ should in principle be modeled as \acp{r.v.}. In this paper, however, perfect \ac{IC} is assumed, thus
%
%\begin{align}\label{residual_interference_expression}
%Q_{vj} = 0 \qquad \forall j \in \{1,2,\dots,\nslot \} \quad \mathrm{and} \quad \forall\, v \in \mathcal C_j \, .
%\end{align}
%
%where $\delta$ is a constant.
%\CS{Maybe some text to backup this assumption. Maybe some text about estimating channel coefficients that vary from slot to slot and connect that to imperfect IC.}
%
Exploiting \eqref{eq:interference_expression}, %
%and \eqref{residual_interference_expression},
after simple manipulation we obtain
\begin{align}
\frac{\powerrv_{ij}}{\noise + \inter_{ij}}
%& = \frac{\powerrv_m}{ \noise + \sum_{i  \in \mathcal{R} \setminus \{m \} } \powerrv_i + \sum_{i \in \mathcal{C}} Q_i} = \frac{\powerrv_m}{ \noise + \sum_{i  \in \mathcal{R} \setminus \{m \} } \powerrv_i + | \mathcal{C} |  \delta  \poweravg }  \\
= \frac{\snrrv_{ij}}{ 1 + \sum_{u  \in \mathcal{R}_j \setminus \{i \} } \snrrv_{uj}} \, .
\end{align}
Hence, in the adopted threshold-based capture model the condition $\frac{\powerrv_{ij}}{\noise + \inter_{ij}} \geq \snrthr$ in \eqref{eq:capture_model} may be recast as
\begin{align}\label{eq:thr}
\frac{\snrrv_{ij}}{ 1 + \sum_{u  \in \mathcal{R}_j \setminus \{ i \} } \snrrv_{uj}} \geq \snrthr \, .
\end{align}

When processing the signal received in some slot $j$, if burst replica $i$ is successfully decoded due to fulfillment of \eqref{eq:thr}, then (i) its contribution of interference is cancelled from slot $j$, and (ii) the contributions of interference of all replicas of the same burst are removed from the corresponding slots.\footnote{We assume that the receiver is able to estimate the channel coefficients required for the removal of the replicas.} Hereafter, we refer to the former part of the \ac{IC} procedure as \emph{intra-slot} \ac{IC} and to the latter as \emph{inter-slot} \ac{IC}. Unlike \ac{SIC} in \ac{IRSA} protocols over a collision channel, which only rely on inter-slot \ac{IC}, \ac{SIC} over a block fading channel with capture takes advantage of intra-slot \ac{IC} to potentially decode burst replicas interfering each other in the same slot. In this respect, it effectively enables \emph{multi-user decoding} in the slot.

Upon reception of a new \ac{MAC} frame, slots are processed sequentially by the receiver. By definition, one \emph{\ac{SIC} iteration} consists of the sequential processing of all $\nslot$ slots. In each slot, intra-slot \ac{IC} is performed repeatedly, until no burst replicas exist for which \eqref{eq:thr} is fulfilled. When all burst replicas in slot $j$ have been successfully decoded, or when intra-slot \ac{IC} in slot $j$ stops prematurely, inter-slot \ac{IC} is performed for all burst replicas successfully decoded in slot $j$ and the receiver proceeds to process slot $j+1$. When all $\nslot$ slots in the \ac{MAC} frame have been processed there are three possible cases: (1) a success is declared if all user packets have been successfully received; (2) a new iteration is started if at least one user packet has been recovered during the last iteration, its replicas removed via inter-slot IC, and there still are slots with interfering burst replicas; (3) a failure is declared if no user packets have been recovered during the last iteration and there still are slots with interfering burst replicas, or if a maximum number of \ac{SIC} iterations has been reached and there still are slots with interfering burst replicas.

Exploiting the graphical representation reviewed in Section~\ref{subsec:graph}, the \ac{SIC} procedure performed at the receiver may be described as a successive removal of graph edges. Whenever a burst replica is successfully decoded in a slot, the corresponding edge is removed from the bipartite graph as well as, due to inter-slot \ac{IC}, all edges connected to the same burst node. A success in decoding the \ac{MAC} frame occurs when all edges are removed from the bipartite graph.
Two important features pertaining to the receiver operation, when casted into the graph terms, should be remarked. The first one is that, due to capture effect, an edge may be removed from the graph when it is connected to a slot node with residual degree larger than one. The second one is that an edge connected to a slot node with residual degree one may not be removed due to poor \ac{SNR}, when  \eqref{eq:thr}, with $\mathcal{R}_{j} \setminus \{i\} = \emptyset$, is not fulfilled. 
\section{Decoding Probabilities}\label{sec:decoding_probs}

Consider the generic slot node $j$ at some point during the decoding of the \ac{MAC} frame and assume it has degree $r$ under the current graph state. This means that $r$ could be the original slot node degree $\slotind_j$ or the residual degree after some inter-slot and intra-slot \ac{IC} processing. Note that, as we assume perfect \ac{IC}, the two cases $r=c_j$ and $r<c_j$ are indistinguishable.

Among the $r$ burst replicas not yet decoded in slot $j$, we choose one randomly and call it the reference burst replica. Moreover, we denote by $D(r)$ the probability that the reference burst replica is decoded starting from the current slot setting and only running intra-slot \ac{IC} within the slot. As we are considering system with $ \snrthr \geq 1$, the threshold based criterion \eqref{eq:thr} can be satisfied only for one single burst replica at a time. Therefore there may potentially be $r$ decoding steps (and $r-1$ intra-slot \ac{IC} steps), in order to decode the reference burst replica. Letting $D (r,t)$ be the probability that the reference bust replica is successfully decoded in step $t$ and not in any step prior to step $t$, we may write
\begin{align}
D( r ) = \sum_{t=1}^{r} D (r,t) \, . \label{eq:D(d,r)}
\end{align}

Now, with a slight abuse of the notation, label the $r$ burst replicas in the slot from $1$ to $r$, arranged such that: (i) the first $t-1$ are arranged by their \acp{SNR} in the descending order (i.e., $\snrrv_{1} \geq \snrrv_{2} \geq \ldots \snrrv_{t-1} ) $, (ii) the rest have \ac{SNR} lower than $\snrrv_{t-1}$ but do not feature any particular \ac{SNR} arrangement among them, (iii) the reference burst is labeled by $t$, i.e., its \ac{SNR} by $\snrrv_{t}$, and (iv) the remaining $ r - t$ bursts are labeled arbitrarily. 
The probability of having at least $t$ successful burst decodings through successive intra-slot \ac{IC} for such an arrangement is
\begin{align}
& \Pr \left\{\frac{\snrrv_{1}}{1 + \sum_{i=2}^{r} \snrrv_{i} } \geq \snrthr, \dots, \frac{\snrrv_{t}}{1 + \sum_{i=t+1}^{r} \snrrv_{i} } \geq \snrthr \right\} \\
&= \frac{1}{ \snravg^r} \int_{0}^{\infty} \mathrm{d} b_{r} \cdots \int_{0}^{\infty}  \mathrm{d} b_{t + 1} \notag \\
& \times \int_{ \snrthr ( 1 + \sum_{i = t+1}^{r} b_i ) }^{\infty} \mathrm{d} b_{t} \, \cdots  \nonumber  \int_{ \snrthr ( 1 + \sum_{i = 2}^{r} b_i )}^{\infty} \mathrm{d} b_{1} e^{- \frac{b_{r}}{\snravg}} \, \cdots \, e^{- \frac{b_{1}}{\snravg}} \\
&=  \frac{e^{-\frac{\snrthr}{\snravg} \sum_{i=0}^{t-1} (1+ \snrthr)^i}}{(1+ \snrthr )^{t ( r - \frac{t+1}{2})} }  = \frac{e^{- \frac{1}{\snravg} ( ( 1 + \snrthr )^t - 1 ) }}{(1+ \snrthr )^{t \left( r - \frac{t+1}{2} \right)}} .
\label{eq:ordered_prob}
\end{align}
%
%With $b^{'}_i=b_i/\snravg$ for $i=1,...,r$ and $\snrthrnorm=\snrthr/\snravg$.
Further, the number of arrangements in which the \ac{SNR} of the reference replica is not among first $(t - 1)$ largest is $\frac{(r-1)!}{(r-t)!}$, where it is assumed that all arrangements are a priory equally likely.
Thus, the probability that the reference burst is decoded exactly in the $t$-th step is
\begin{align}
\label{eq:D(r|t)}
D (r,t) = \frac{(r-1)!}{(r-t)!}   \frac{e^{ - \frac{1}{\snravg} ((1 + \snrthr)^t - 1) }}{(1+ \snrthr )^{t \left( r - \frac{t+1}{2} \right)}} , \; 1 \leq t \leq r.
\end{align}

We conclude this section by noting that  $D ( 1 ) =  e^{ - \frac{\snrthr}{\snravg} } \leq 1$, i.e., a slot of degree 1 is decodable with probability that may be less than 1 and that depends on the ratio of the capture threshold and the expected SNR.
Again, this holds both for slots whose original degree was 1 and for slots whose degree was reduced to 1 via IC, as these two cases are indistinguishable when the IC is perfect.

\section{Density Evolution Analysis and Decoding Threshold Definition}\label{sec:DE}

\begin{figure}[tb]
\begin{center}
\includegraphics[width=\columnwidth,draft=false]{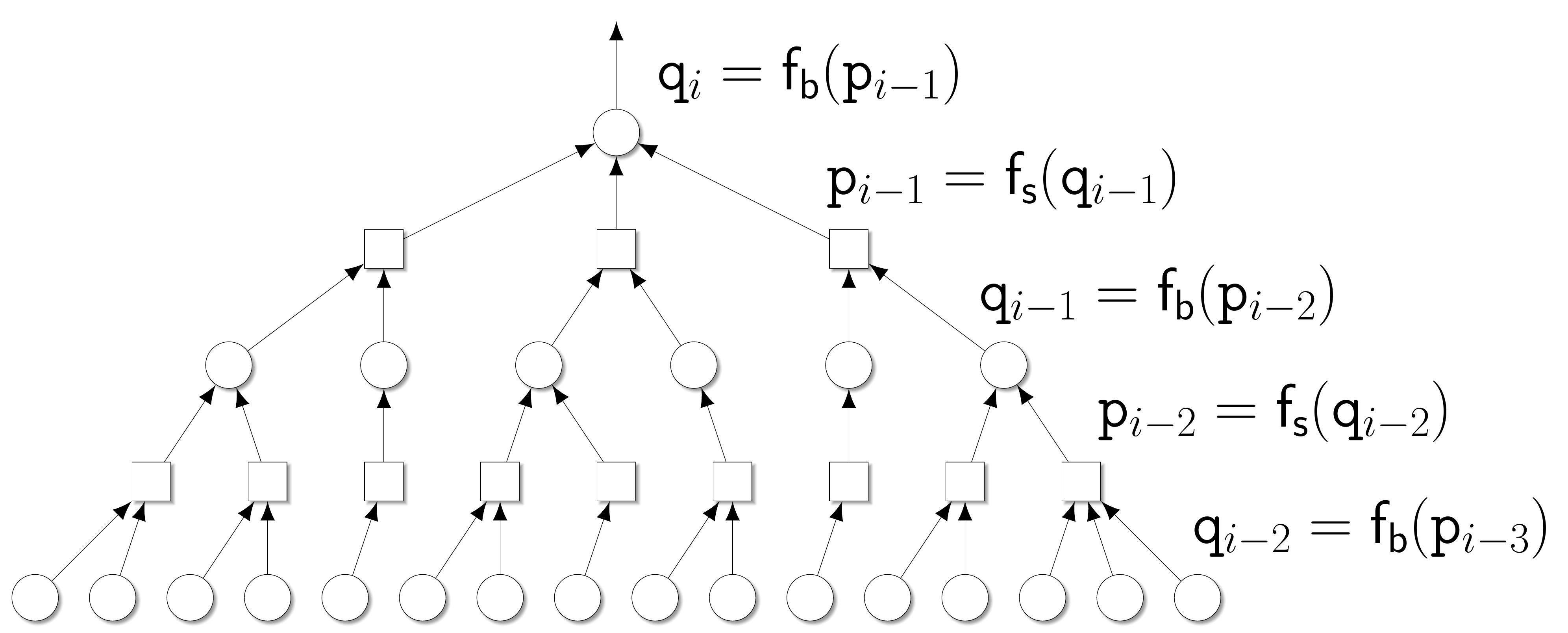}
\end{center}
\caption{Tree representation of the MAC frame.}\label{fig:and_or_tree}
\end{figure}

In this section, we apply the technique of \ac{DE} in order to evaluate asymptotic performance of the proposed technique, i.e., when $\nuser \rightarrow \infty$ and $ \nslot \propto \nuser$.
For this purpose, we unfold the graph representation of the MAC frame (Fig.~\ref{fig:graph}) into a tree, choosing a random burst node as its root, as depicted in Fig.~\ref{fig:and_or_tree}.
The evaluation is performed in terms of probabilities that erasure messages are exchanged over the edges of the graph, where the erasure message denotes that the associated burst is not decoded.\footnote{For a more detailed introduction to the \ac{DE}, we refer the interested reader to \cite{RU2007}.}
The message exchanges are modeled as successive (i.e., iterative) process, corresponding to the decoding algorithm described in Section~\ref{subsec:receiver} in the asymptotic case, when the lengths of the loops in the graph tends to infinity.
Specifically, the $i$-th iteration consists of the update of the probability $\q_i$ that an edge carries an erasure message from a burst node to a slot node, followed by the update of the probability $\p_i$ that an edge carries an erasure message from a slot node to a burst node. These probabilities are averaged over all edges in the graph. We proceed by outlining the details.

The probability that an edge carries an erasure message from burst nodes to slot nodes in $i$-th iteration is
\begin{align}
\label{eq:final_q}
\q_i = \sum_{\ind=1}^{\dmax} \ld_\ind \, \q_i^{(\ind)}  = \sum_{\ind=1}^{\dmax} \ld_\ind \, \p_{i-1}^{\ind-1} =: \exitb(\p_{i-1})
\end{align}
where $\ld_\ind$ is the probability that an edge is connected to a burst node of degree $\ind$ (see Section~\ref{subsec:graph}) and $\q_i^{(\ind)}$ is the probability that an edge carries an erasure message given that it is connected to a burst node of degree $\ind$. In the second equality we used the fact that the an outgoing message from a burst node carries an erasure only if all incoming edges carry an erasure, i.e., $ \q_i^{(\ind)} = \p_{i-1}^{\ind-1}$.

Similarly, the probability that an edge carries an erasure message from SNs to BNs in $i$-th iteration is
\begin{align}
\label{p_i_avg}
\p_i & = \sum_{\slotind=1}^{+\infty} \rd_\slotind \, \p_i^{(\slotind)}
\end{align}
where $\rd_\slotind$ is the probability that an edge is connected to a slot node of degree $\slotind$, and where $\p_i^{(\slotind)}$ is the probability an edge carries an erasure message given that it is connected to a slot node of degree $\slotind$.
This probability may be expressed as
\begin{align}
\label{p_i_d}
\p_i^{(\slotind)} = 1 - \sum_{r=1}^{\slotind} D( r ) { \slotind - 1 \choose r - 1} \q_{i}^{r-1} ( 1 - \q_{i} )^{\slotind-r},
\end{align}
where summation is done over all possible values of the reduced degree $r$, i.e., $ 1 \leq r \leq \slotind$, and where the term ${ \slotind - 1 \choose r - 1} \q_{i}^{r-1} ( 1 - \q_{i} )^{\slotind - r}$ corresponds to the probability that the degree of the slot node is reduced to $r$ and $D (r )$ is the probability that the burst corresponding to the outgoing edge is decoded when the (reduced) degree of the slot node is $r$.\footnote{As in the asymptotic case the loops in the graph are assumed to be of infinite length, such that the tree representation in Fig.~\ref{fig:and_or_tree} holds, the reduction of the slot degree happens only via inter-slot IC, which is implicitly assumed in the term ${ \slotind - 1 \choose r - 1} \q_{i}^{r-1} ( 1 - \q_{i} )^{\slotind - r}$. On the other hand, $D (r)$ expresses the probability that an outgoing edge from the slot node is decoded using intra-slot IC (see Section~\ref{sec:decoding_probs}). In other words, inter- and intra-slot IC are in the asymptotic evaluation separated over \ac{DE} iterations.}

Combining \eqref{p_i_d} and the expression for the edge-oriented slot-node degree distribution (see Section~\ref{subsec:graph}) into \eqref{p_i_avg} yields
\begin{equation}
\p_i = 1 - e^{ - \frac{\load }{ \rate } } \sum_{\slotind=1}^{\infty} \left( \frac{\load}{ \rate} \right)^{\slotind - 1} \sum_{r=1}^{\slotind} \frac { D ( r ) }{ ( r - 1 ) ! } \q_i^{r-1} ( 1 - \q_i )^{\slotind-r}.\label{eq:p_i}
\end{equation}
It can be shown that in case of perfect IC, \eqref{eq:p_i} becomes
\begin{align}
\p_i & = 1 - e^{-\frac{\load}{\rate}} \sum_{r = 1}^{+\infty} \frac{D ( r )}{ ( r - 1 )! } \left( \frac{\load}{\rate} \q_i \right)^{ r - 1 } \sum_{\slotind = 0}^{+\infty} \frac{\left( \frac{\load}{\rate} ( 1 - \q_i ) \right)^{ \slotind } }{\slotind !}\\
& = 1 - e^{-\frac{\load}{\rate} \q_i } \sum_{r=1}^{+\infty} \frac{D( r )}{ ( r - 1 )!} \left( \frac{\load}{\rate} \q_i \right)^{r - 1 }, \label{eq:p_i_1}
\end{align}
where $ D ( r ) = \sum_{t =1}^{r} D ( r ,t )$, see \eqref{eq:D(r|t)}.
%\begin{align}
%\p = \exits(\q) :&= 1 - \sum_{d=1}^{\infty} \rd_d \sum_{r=1}^{d} D(d,r) { d - 1 \choose r - 1} \q^{r-1} ( 1 - \q)^{d-r} \label{eq:DE:exitq}
%\end{align}
Further, defining $\zt=(1+\snrthr)^t$, \eqref{eq:p_i_1} becomes
\allowdisplaybreaks
\begingroup
\begin{align}
\label{eq:final_p}
\p_i & = 1 - e^{-\frac{\load}{\rate} \q_i } \sum_{r=1}^{+\infty} \left( \frac{\load}{\rate} \q_i \right)^{r - 1 } \sum_{t = 1}^{r} \frac{e^{ - \frac{1}{\snravg} ( \zt - 1 )}}{( r - t )! \zt^{ \left(r - \frac{t+1}{2}\right) }} \notag \\
& = 1 - e^{-\frac{\load}{\rate} \q_i } \sum_{t=1}^{+\infty} \frac{ \left( \frac{\load}{\rate} \q_i \right)^{t-1} }{ \zt^{\left(\frac{ t - 1 }{ 2 }\right)}} e^{ - \frac{1}{\snravg} ( \zt - 1 )} \sum_{r = 0 }^{+\infty} \frac{ \left( \frac{\load}{\rate} \q_i \right)^r }{r ! \zt^r } \notag \\
%& = 1 - e^{-\frac{\load}{\rate} \q_i } \sum_{t=1}^{+\infty} \frac{ \left( \frac{\load}{\rate} \q_i \right)^{t-1} }{ \zt^{\left(\frac{ t - 1 }{ 2 }\right)}} e^{ - \frac{1}{\snravg} ( \zt - 1 )}  e^{ \frac{\load}{\rate} \q_i \zt^{-1} } \notag \\
& = 1-  \sum_{t=1}^{+\infty} \frac{ \left( \frac{\load}{\rate} \q_i \right)^{t-1} }{ \zt^{\left(\frac{ t - 1 }{ 2 }\right)}} e^{ - ( \zt - 1 ) \left (\frac{1}{\snravg} + \frac{\frac{\load}{\rate} \q_i} {\zt }\right)  } =: \exits(\q_i) \, .
\end{align}
\endgroup

%After some manipulation, it can be shown that \CS{to check}
%\begin{align}
%\p_i = 1 - \sum_{r=1}^{+ \infty} \left( \frac{ \frac{\bar{n}}{k} G \q_i }{( 1 + \snrthr) ^{\frac{r}{2}} }  \right)^{k-1} e^{ \left( ( 1 + \snrthr )^k - 1 \right) \left( \frac{1}{\snravg} + \frac{\frac{\bar{n}}{k} G \q_i}{( 1 + \snrthr )^k} \right)}
%\end{align}

A \ac{DE} recursion is obtained combining \eqref{eq:final_q} with \eqref{eq:final_p}, consisting of one recursion for $\q_i$ and one for $\p_i$. In the former case, the recursion assumes the form $\q_i = (\exitb \circ \exits) (\q_{i-1})$ for $i \geq 1$, with initial value $\q_0 = 1$. In the latter case, it assumes the form $\p_i = (\exits \circ \exitb) (\p_{i-1})$ for $i \geq 1$, with initial value $\p_0 = \exits(1)$. Note that the \ac{DE} recursion for $\p_i$ allows expressing the asymptotic \ac{PLR} of an \ac{IRSA} scheme in a very simple way. More specifically, let $\p_{\infty} (\load, \BNdegDist, \snravg,\snrthr) = \lim_{i \rightarrow \infty} \p_i$ be the limit of the \ac{DE} recursion, where we have explicitly indicated that the limit depends on the system load, on the burst node degree distribution, on the average \ac{SNR}, and on the threshold for successful intra-slot decoding. Since $\left[ \p_{\infty}(\load, \BNdegDist, \snravg, \snrthr) \right]^{\ind}$ represents the probability that a user packet associated with a burst node of degree $d$ is not successfully received at the end of the decoding process, the asymptotic \ac{PLR} is given by
\begin{align}
\PLR(\load,\BNdegDist,\snravg, \snrthr) = \sum_{\ind=2}^{\dmax} \Ld_\ind \left[ \p_{\infty} (\load, \BNdegDist, \snravg, \snrthr) \right]^{\ind} \, .
\end{align}

Next, we introduce the concept of \emph{asymptotic decoding threshold} for an \ac{IRSA} scheme over the considered block fading channel model and under the decoding algorithm described in Section~\ref{subsec:receiver}. Let $\PLRtarget$ be a target \ac{PLR}. Then, the asymptotic decoding threshold, denoted by $\load^{\star} = \load^{\star}(\BNdegDist, \snravg, \snrthr, \PLRtarget)$, is defined as the supremum system load value for which the target \ac{PLR} is achieved in the asymptotic setting:
\begin{align}
\load^{\star} = \sup_{\load \geq 0} \{ \load : \PLR(\load,\BNdegDist,\snravg, \snrthr) < \PLRtarget \}.
\end{align}
\section{Numerical Results}\label{sec:Performance}

Table~\ref{tab:Distribution} shows some degree distributions designed combining the \ac{DE} analysis developed in Section~\ref{sec:DE} with the differential evolution optimization algorithm proposed in \cite{diffEvol1997}. For each design we set $\PLRtarget=10^{-2}$, $\snravg=20$ dB, and $\snrthr=3$ dB, and we constrained the optimization algorithm to find the distribution $\BNdegDist$ with the largest threshold $\load^{\star}$ subject to a given average degree\footnote{A constraint on the average degree $\davg$ can be turned into a constraint on the rate $\rate$ as there is a direct relation between the two; see also equation \eqref{eq:rate}.} $\davg$ and maximum degree $\dmax=16$.

For all chosen average degrees, the $\load^{\star}$ threshold of the optimized distribution largely exceeds the value $1\,\mathrm{[packet/slot]}$, the theoretical limit under a collision channel model. In general, the higher the average degree $\davg$, the larger is the load threshold $\load^{\star}$. However, as $\davg$ increases, more complex burst node distributions are obtained. For instance, under a $\davg=4$ constraint, the maximum degree is $\dmax=16$ (i.e., a user may transmit up to $16$ copies of its packet); when reducing $\davg$, the optimization converges to degree distributions with a lower maximum degree and degree-$2$ nodes become increasingly dominant.

To assess the effectiveness of the proposed design approach, tailored to the block fading channel with capture, we optimized a distribution $\Ld_5(x)$ using the \ac{DE} recursion over the collision channel \cite{Liva11:IRSA} and again constraining the optimization to $\davg=4$ and $\dmax=16$. As from Table~\ref{tab:Distribution}, due to the mismatched channel model, a $7\%$ loss in terms of $\load^{\star}$ threshold is observed w.r.t. the distribution $\Ld_1(x)$ that fulfills the same constraints but was obtained with the \ac{DE} developed in this paper.

\begin{table}[tb]

\caption{Optimized user node degree distribution and corresponding threshold  $\thr$ for $\PLRtarget=10^{-2}$.}\label{tab:Distribution}

\begin{center}
\begin{tabularx}{\columnwidth}{c|>{\centering}X|c}
\hline\hline
\rule{0pt}{3ex}
$\davg$ & Distribution $\Ld(x)$ & $\thr$ \\[1mm]
\hline
\rule{0pt}{2.5ex}
$4$ & $\Ld_1(x)=0.59 x^2 + 0.27 x^3 + 0.02 x^5 + 0.12 x^{16}$ & $1.863$ \\
\rule{0pt}{0ex}
$3$ & $\Ld_2(x)=0.61 x^2 + 0.25 x^3 + 0.03 x^6 + 0.02 x^7 + 0.07 x^8 + 0.02 x^{10}$ & $1.820$ \\
$2.5$ & $\Ld_3(x)=0.66 x^2 + 0.16 x^3 + 0.18 x^4$ & $1.703$ \\
$2.25$ & $\Ld_4(x)=0.65 x^2 + 0.33 x^3 + 0.02 x^4$ & $1.644$ \\
$4$ & $\Ld_5(x)=0.49 x^2 + 0.25 x^3 + 0.01 x^4 + 0.03 x^5 + 0.13 x^6 + 0.01 x^{13} + 0.02 x^{14} + 0.06 x^{16}$ & $1.734$ \\
\hline\hline
\end{tabularx}
\end{center}
\end{table}

\begin{figure}[tb]
\begin{center}
\includegraphics[width=0.8\columnwidth,draft=false]{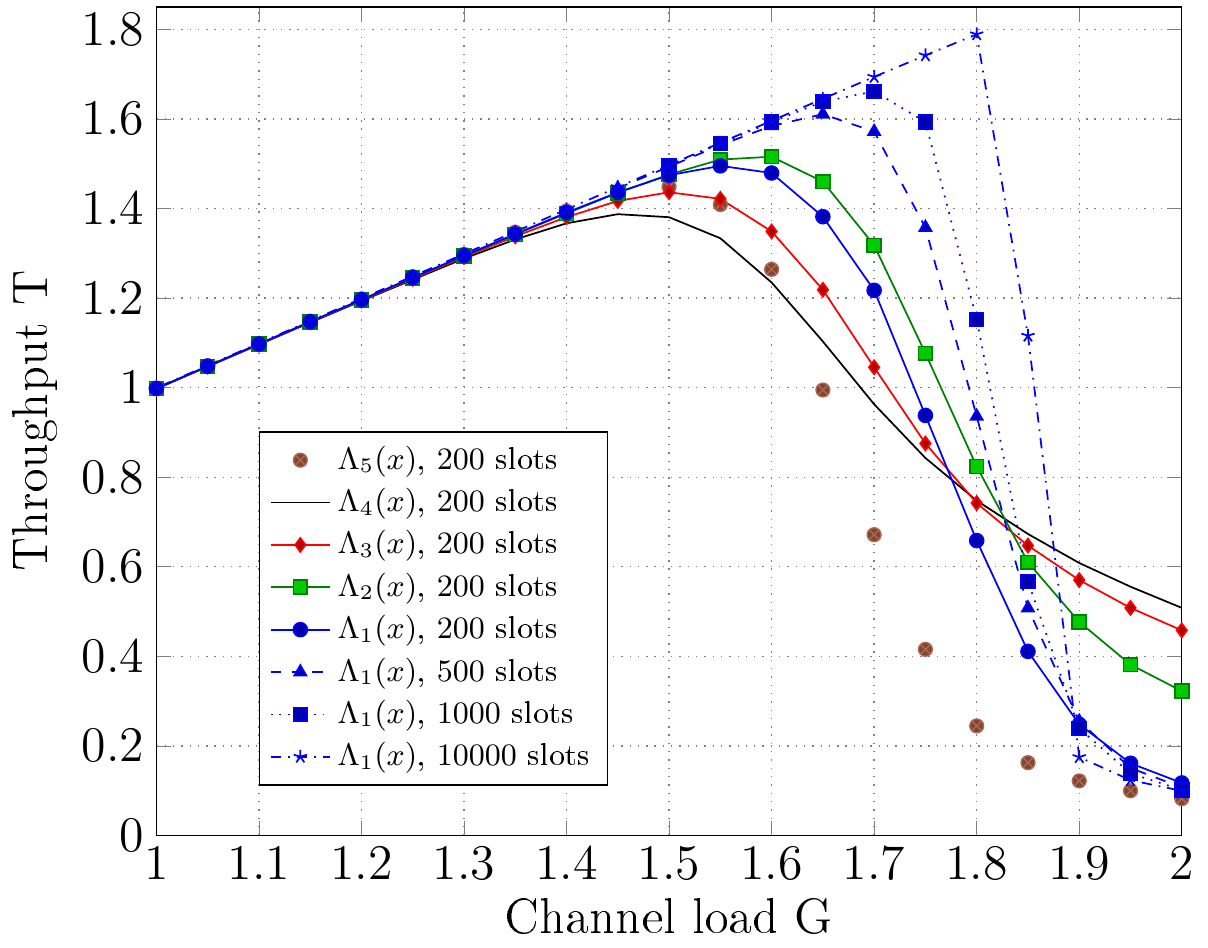}
\end{center}
\caption{Throughput values achieved by the burst node distributions in Table~\ref{tab:Distribution}, versus the channel load. Various frame sizes $\nslot$, $\snravg=20$ dB, $\snrthr=3$ dB.}\label{fig:thr}
\end{figure}

We tested the optimized distributions performance through Monte Carlo simulations for finite frame lengths, setting $\nslot=200$ (unless otherwise stated), $\snravg=20$ dB, $\snrthr=3$ dB, and the maximum number of \ac{IC} iterations to $20$. Fig.~\ref{fig:thr} illustrates the throughput $\tp$, defined as the average number of successfully decoded packets per slot, versus the channel load. For relatively short frames of $\nslot=200$, all distributions exhibit peak throughputs exceeding $1\,\mathrm{[packet/slot]}$. The distribution $\Ld_2(x)$ is the one achieving the highest throughput of $1.52\, \mathrm{[packet/slot]}$ although $\Ld_1(x)$ is the distribution with the highest threshold $\thr$. It is important to recall that, the threshold is computed for a target \ac{PLR} of $\PLRtarget=10^{-2}$ while considering finite frame lengths, the threshold effect on the \ac{PLR} tends to vanish and a more graceful degradation of the \ac{PLR} curve as the channel load increases is expected. Moreover, as there are user nodes transmitting as high as $16$ replicas, the distribution $\Ld_1(x)$ is more penalized for short frame sizes w.r.t. to $\Ld_2(x)$ where at most $10$ replicas per user node are sent.\footnote{Results for frame size of $500$ slots, not presented in the figures, show that the peak throughput for $\Ld_1(x)$ is $1.61$ while for $\Ld_2(x)$ is $1.60$.} This effect, coupled with the fact that the threshold $\thr$ of $\Ld_1(x)$ is only slightly better than the one of $\Ld_2(x)$, explains the peak throughput behavior. To investigate the benefit of larger frames, we selected $\Ld_1(x)$ and we increased the frame size up to $10000$ slots. As expected, the peak throughput is greatly improved from $1.49$ to $1.79\, \mathrm{[packet/slot]}$, i.e., $20\%$ of gain.
\begin{figure}[tb]
\begin{center}
\includegraphics[width=0.8\columnwidth,draft=false]{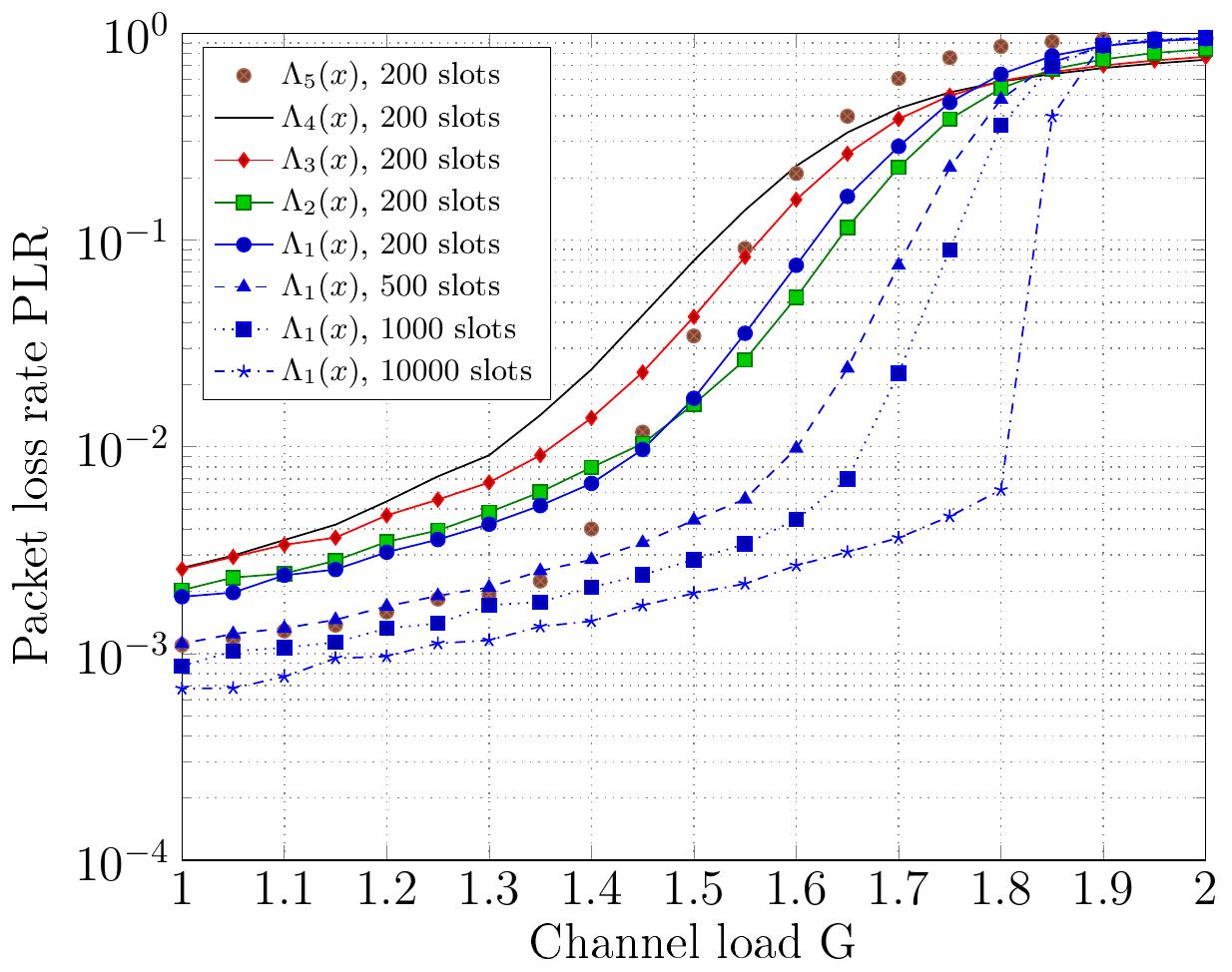}
\end{center}
\caption{\ac{PLR} values achieved by the burst node distributions in Table~\ref{tab:Distribution}, versus the channel load. Various frame sizes $\nslot$, $\snravg=20$ dB, $\snrthr=3$ dB.}\label{fig:plr}
\end{figure}
The \ac{PLR} performance is illustrated in Fig. \ref{fig:plr}. Coherently with the optimization results, the $\Ld_1(x)$ distribution achieves $\PLR=10^{-2}$ for values of the channel load slightly larger than the ones required by $\Ld_2(x)$. Indeed, the steeper \ac{PLR} curve of $\Ld_1(x)$ is the reason for the slightly larger peak throughput of $\Ld_2(x)$ observed in Fig. \ref{fig:thr}.
%As the average degree is increased from $\Ld_4(x)$ to $\Ld_1(x)$ a beneficial impact below the target \ac{PLR} can be observed.
Finally, as expected, an increase of the number of slots per frame yields an increase of the channel load for which the target \ac{PLR} is achieved.

%\begin{figure}[tb]
%\begin{center}
%\includegraphics[width=0.7\columnwidth,draft=false]{./figures/Bound/Bound.pdf}
%\end{center}
%\caption{Bound.}\label{fig:bound}
%\end{figure} 

\section{Conclusions}\label{sec:Conclusions}

The asymptotic analysis of \ac{IRSA} access schemes, assuming both a Rayleigh block fading channel and capture effect, was presented in the paper. We derived the decoding probability of a burst replica in presence of intra-slot \ac{IC}. The \ac{DE} analysis is modified considering the Rayleigh block fading channel model, and the user/slot nodes updates of the iterative procedure are explicitly derived. Due to the presence of fading, the optimization procedure target has been modified as well. The distribution able to achieve the highest channel load value without exceeding a properly defined \ac{PLR} target is selected. We designed some degree distributions with different values of average degree. Remarkably, all of them present a load threshold that guarantees \ac{PLR} below $10^{-2}$ for values well above $1\,\mathrm{[packet/slot]}$. The best distribution exceeds $1.8\,\mathrm{[packet/slot]}$. The derived distributions were shown to perform well also for finite frame durations. In a frame with $200$ slots, the peak throughput exceeds $1.5\,\mathrm{[packet/slot]}$ and up to $1.45\,\mathrm{[packet/slot]}$ the \ac{PLR} remains below $10^{-2}$. 

\section*{Acknowledgement}

The authors would like to thank Dr. Gianluigi Liva (German Aerospace Center) and Prof. Marco Chiani (University of Bologna)for fruitful discussions.
The work of \v C. Stefanovi\' c was supported by the Danish Council for Independent Re-search under grant no. DFF-4005-00281.
The work of E. Paolini was supported by ESA/ESTEC under Contract no. 4000118331/16/UK/ND ``SCAT''.

\end{document}